\documentstyle[aps,preprint]{revtex}
\begin{document}
\draft
\title{A Gaussian Theory of Superfluid$-$Bose-Glass Phase Transition}
\author{Pornthep Nisamaneephong$^{1}$,
Lizeng Zhang$^{2,3}$, \\ and Michael Ma$^{1,4}$}
\address{ $^{1}$Department of Physics, University of Cincinnati \\
Cincinnati, Ohio 45221-0011 }
\address{ $^{2}$ Department of Physics and Astronomy, The University
of Tennessee\\ Knoxville, Tennessee 37996-1200 }
\address{ $^{3}$ Solid State Division, Oak Ridge National
Laboratory\\ Oak Ridge, Tennessee 37831}
\address{ $^{4}$ Department of Physics, Hong Kong University of
Science and Technology \\ Clearwater Bay, Kowloon, Hong Kong}
\maketitle
\begin{abstract}
We show that gaussian quantum fluctuations, even if infinitesimal,
are sufficient to destroy the superfluidity of a disordered boson
system in 1D and 2D. The critical disorder is thus finite no matter
how small the repulsion is between particles. Within the gaussian
approximation, we study the nature of the elementary excitations,
including their density of states and mobility edge transition. We
give the gaussian exponent $\eta$ at criticality in 1D and show that
its ratio to $\eta$ of the pure system is universal.
\end{abstract}

In the presence of disorder, (repulsive) interacting bosons can
undergo a transition from the superfluid (SF) phase into an insulating
Bose-glass (BG) phase \cite{MHL} - \cite{MC}.  This transition
is intrinsically quantum in nature in that no amount of disorder
will destroy the superfluidity without invoking the non-commutativity
of density $\rho$ and phase $\phi$. Hence, the usual
saddle point or Hartree solution is always long-range ordered, and
corresponds to a non-uniform condensate. Given that, it is of interest
to investigate what `minimal' quantum effects are necessary to give a
transition. In terms of going beyond the saddle point approximation,
these effects can be characterized as gaussian, non-linear, topological
(as in vortices) etc. In effect, one is asking `what drives the
transition', even if the true universality class of the transition may
require quantum fluctuations beyond the `minimal' ones. To clarify
this perspective, consider the 2D classical XY model as an analogy.
There, the true long range order (LRO) is
destroyed at any finite temperature by spin waves, even though
(bound) vortices do renormalize the exponent $\eta$ (universality
class). That is, spin wave alone can explain why the low temperature
phase has algebraically decaying correlations. On the other hand,
vortices must be invoked to explain the Kosterlitz-Thouless
transition \cite{PB}.

In this article we show that gaussian fluctuations, even
if infinitesimal, are sufficient to destroy superfluidity
in 1D and 2D at finite disorder. The model we use is the
hard-core boson model with on-site disorder, which is
equivalent to the spin-1/2 XY magnet with a transverse random field
\cite{MHL,ZM1,ZM2}.  Written in a rotated frame for later
convenience, the Hamiltonian is \cite{LZ}:
\begin{equation}\label{XYh}
{\cal H} = -J\sum_{<i,j>}(S_{i}^{z}S_{j}^{z} + S_{i}^{x}S_{j}^{x})
- \sum_{j} h_{j} S_{j}^{y} \;\; ,
\end{equation}
where the random field $\{ h_{j}\}$ is given by independent
gaussian distribution function $P(h_{j})$
with width $h$. We pick this model because while it contains
the features believed to be essential for the SF-BG transition, it has a
simple classical solution and a `built-in' parameter to systematically
investigate quantum fluctuations. It is also of interest as a
disordered quantum spin system \cite{ZM1} - \cite{LZ}.
Our study reported in this paper will be focused on the $T=0$ case.

The off-diagonal LRO of the boson system is related to the magnetic
LRO in the $x-z$ plane. The classical solution for this model
corresponds to treating the spins as classical vectors. In terms of the
bosons, one is dealing with variational wavefunction of the form
$\prod_{j} (u_{j} + v_{j}b^{+}_{j})|0\rangle$, which when projected
into states of definite $N$ are Jastrow wavefunctions given by
Gutzwiller projection of condensate wavefunction
$(\sum_{j}\frac{v_{j}}{u_{j}}b^{\dagger}_{j})^{N}|0\rangle$. In Ref.
\cite{MHL}, it was shown in the spin model that provided there is no
gap in the spectrum, the classical ground state is always ordered.
However, it was also seen that the ground state is not long
range ordered with strong disorder when the quantum
(specifically $S=1/2$) nature of the spin operators is
taking into account.  Following the
motivation discussed in the previous paragraph,
it is thus of interest to investigate if the destruction of LRO can be
achieved by gaussian quantal effects. This can be studied by means
of a spin wave analysis \cite{LZ}.

Within such an approach, the first question is what would be the
signature of destruction of the LRO \cite{HM}. Since the spin-wave analysis
is an expansion about the ordered state, this destruction is indicated
by an instability. Possible scenarios are 1) a diverging fluctuation in
the order parameter, 2) negative excitation energies, or 3) complex
excitation energies (e.g., Bogoliubov's solution to bosons with attractive
interactions). In the pure case, scenario 1) is observed in 1D. Exact
solution for $S = 1/2$ \cite{LSM} and general understanding of 1D
spin systems indicates that this diverging fluctuations destabilize
the LRO and the ground state has algebraically decaying correlation
functions.

We now derive the spin-wave Hamiltonian \cite{LZ}. First we
generalize Hamiltonian (\ref{XYh}) to arbitrary spin $S$ by rescaling
$J \rightarrow J/S^{2}$ and $h_{j} \rightarrow h_{j}/S$. In the
infinite $S$ limit, the spins behave classically. Taking the $z$-axis as
the ordering axis, the spin on site $j$ lies on the $y-z$ plane at angle
$\theta_{j}$ from the $z$-axis, with $\{\theta_{j} \}$ given
self-consistently by
\begin{equation}\label{ceom}
\sin\theta_{j}J\sum_{<j'>}\cos\theta_{j'} = h_{j}\cos\theta_{j} \;\; ,
\end{equation}
where $<j'>$ indicates nearest neighbors of the site $j$.
The statement that LRO persists to all order is revealed by
the solution to (\ref{ceom}) having all $\cos\theta_{j} \neq 0$
no matter what value of $h$ is. A local rotation about the
$x$-axis is performed so that the spin points along the new
$z$-axis. The usual Holstein-Primakoff transformation \cite{HP}
of the spins into boson operators can now be defined in the
rotated frame. To order $1/S$, one arrives at a quadratic
Hamiltonian for the bosons \cite{LZ}:
\begin{equation}\label{hrrs}
{\cal H} = - J\sum_{<i,j>}\cos\theta_{i}\cos\theta_{j} -
\sum_{j}h_{j}\sin\theta_{j} - \frac{1}{2S}\sum_{<i,j>} \left
(J_{ij}a_{i}^{\dagger}a_{j}+K_{ij}a_{i}a_{j}+H.c. \right )+{\cal
O}(\frac{1}{S^{3/2}}) \;\;,
\end{equation}
where $J_{ij}=J(1+\sin\theta_{i}\sin\theta_{j})+
\frac{h_{j}}{\sin\theta_{j}}\delta_{ij}$ and
$K_{ij}=J(1- \sin\theta_{i}\sin\theta_{j})$,
which describes gaussian fluctuations of strength $1/S$
about the classical ground state. In Ref. \cite{LZ}, (\ref{hrrs}) is
studied perturbatively for weak disorder. In this paper, we will
diagonalize (\ref{hrrs}) numerically on finite-sized lattices, and will
not limit ourselves to weak disorder. This will enable us to study the
destruction of LRO. (\ref{hrrs}) is formally diagonalized by a
Bogoliubov transformation \cite{LLP}:
\begin{equation}
a_{j} = \sum_{\alpha}( u_{j\alpha}\gamma_{\alpha} +
v_{j\alpha}\gamma_{\alpha}^{\dagger}) \;\;,
\end{equation}
where $\alpha$ is the eigenstate index. We have taken the $u$'s
and $v$'s to be real. The $\gamma$'s are boson operators if
\begin{equation}\label{norm}
\sum_{j}( u_{j\alpha}u_{j\alpha'} -
v_{j\alpha}v_{j\alpha'}) = \delta_{\alpha \alpha'} \;\;,
\end{equation}
and we seek the solution
\begin{equation}
{\cal H}_{SW}=E^{0}+\sum_{\alpha}{\omega}_{\alpha}
\gamma_{\alpha}^{\dagger}\gamma_{\alpha} \;\; ,
\end{equation}
which implies the Bogoliubov equations for $u$'s and $v$'s,
\begin{eqnarray}\label{Beqn}
\omega u_{j\alpha} =-\sum_{<j'>}(J_{jj'}u_{j'\alpha} +
K_{jj'}v_{j'\alpha}) \;\;,\nonumber \\
\omega v_{j\alpha} =\sum_{<j'>}(K_{jj'}u_{j'\alpha} +
J_{jj'}v_{j'\alpha}) \;\;,
\end{eqnarray}
to be `normalized' by the condition (\ref{norm}). For $N$ sites, this is
a $2N \times 2N$ matrix equations with $2N$ eigenstates. Note that
for a given solution with eigenvalue $\omega$, there is the
complimentary solution $u \leftrightarrow v$, with eigenvalue
$-\omega$. However, only one of these can be consistent with
(\ref{norm}), and the other is unphysical, leaving us with $N$
physical solutions. The Goldstone mode, corresponding to uniform
spin rotation about the $y$-axis in (\ref{XYh}), is given by
$u_i=v_i\propto \cos\theta_{i}$.

We investigate LRO instability in 1D and 2D. Calculations
in 1D are done on lattices of size 50 - 120, averaging over
500 configurations for each value of $\Delta \equiv J/h$,
and in 2D on $6\times 6$ to $11 \times 11$ lattices
averaging over 200 configurations. Instability criteria 2) and 3) are
not observed, leaving 1), a diverging fluctuation in the order
parameter as the sole possibility. Within the spin wave
approximation as formulated, the relevant quantity is
\begin{equation}
\delta m =\frac{1}{N}\sum_{j}\cos\theta_{j} \delta\langle
S_{j}^{z}\rangle = \frac{1}{N}\sum_{j} \sum_{\alpha \neq
0}\cos\theta_{j}v_{j\alpha}^{2} = \int d\omega
N(\omega)v^{2}(\omega) \;\; ,
\end{equation}
where $N(\omega)=\frac{1}{N}\sum_{\alpha}\delta(\omega-
\omega_{\alpha})$ is the density of states (DOS).

As remarked earlier, in 1D $\delta m$ diverges as $N \rightarrow
\infty$ even without disorder, so it seems criteria 1) is inapplicable.
However, more precisely, $\delta m \propto \ln N$, and we view this
as an indication for an algebraic LRO (replacing the true LRO, see later),
hence the ground state is still a superfluid. Thus, we argue that
the transition is marked by $\delta m$ diverging faster than $\ln N$.
This is in fact seen in our calculation, and is shown in Fig. 1, with the
critical value of $\Delta=\Delta_c \approx 0.6$ in the present model.
The transition occurs thus at finite disorder. Since the spin wave
approximation is correct in the large $S$ limit, there is a
discontinuity between $S \rightarrow \infty$ and $S = \infty$. In 2D,
$\delta m/m$ is finite in the pure case as $N \rightarrow \infty$,
which we take to mean the LRO is stable, and is consistent with the
exact result for $S =1/2$ \cite{KLS}. Fig. 2 shows $\delta m/m$ vs.
ln$N$ for different values, and we see that there is a transition
between $\Delta = 0.1$ and $\Delta = 0.08$.

Thus, already in the gaussian approximation, in contrast to the
classical case, there is a transition from an algebraic in 1D and a true
long range ordered in 2D superfluid phase to a disordered phase.
Since in a gaussian theory, the ground state is just the classical state
modified by the zero point motion of the the excitations, it is of
interest to ask whether the transition is due to a change in the
DOS ($N(\omega)$) or the nature of the excitations
($v^{2}(\omega)$) or both. In the pure case, $v^{2}(\omega) \propto
\frac{1}{\omega}$ for small $\omega$, while $N(\omega) \propto
\omega^{d-1}$ for small $\omega$. For the infinitely strong disorder
($J=0$) case, the excitations are single spin flips, with excitation
energies $|h_{j}|$. Hence $N(\omega)$ is simply given by the
distribution of $h_{j}$, and is finite at low energies. It seems
reasonable to expect therefore $N_{0}=N(\omega \rightarrow 0)$ is
finite in 1D for all $\Delta$, and the transition must be due to
$v^{2}(\omega)$ diverging faster than $1/\omega$. This picture is
confirmed by our numerical calculations and Fig. 3 is shown for DOS
in 1D. While there is some ambiguity in deciding $N_{0}$ for infinite
system from finite-size calculation, we have checked to see that the
scaling of $N_{0}$ with $N$ is in fact consistent with a non-zero DOS
at zero energy. For $\Delta < \Delta_{c}$, $\delta m \propto
N^{\theta}$, with $\theta = \theta(\Delta)$. We find that this
exponent is in agreement with the exponent of $v^{2}(\omega)
\propto \frac{1}{\omega^{\delta}}$, with $\delta = 1+\theta$, again
indicating $N_{0}$ finite. In 2D, the ordered phase should be
characterized by $N(\omega) \propto \omega$ and the disordered
phase by $N_{0}$ finite. Our results are consistent with this. For
$\omega \geq 0.1$, $N(\omega)$ is linear in $\omega$, with the
slope increasing with decreasing $\Delta$. For $\Delta \leq 0.08$,
$N_{0}$ is finite. Unfortunately, we cannot say for certain whether
the DOS transition exactly occurs at the order parameter transition
due to the inability of pinpointing $\Delta_{c}$ (The popular
finite-size scaling method for locating critical point is not
applicable here since there is no scale invariance). We hope
to clear up this point in a later publication.

The low-energy excitations calculated here
 are particularly significant in the SF
phase, as they are approximations to the collective modes (phonons).
These excitations can be extended or localized. It is of interest to ask
if their localization transition is related to the `localization' of the
ground state. It is also of interest by itself as an Anderson
localization problem of the eigenstates of (3). Since the zero-mode
corresponds to uniform phase rotation, it must be extended. One thus
expects that possibly for a given $\Delta$, a transition from extended
to localized states with increasing energy at a mobility edge energy
$E_{c}$, and perhaps $E_{c} \rightarrow 0$ as $\Delta \rightarrow
\Delta_{c+}$. The quantity we calculate as a measure of localization is
the inverse participation ratio $p$, which we assume for localized
states scale as the inverse localization length $\xi^{-1}$, and is zero
for extended states in infinite systems. $E_{c}$ is the energy where
$p$ first vanishes. However, for finite size, $p$ scales as the greater
of $\xi^{-1}$, $L^{-1}$. Hence, at low energy, where $\xi > L$, $p(E)$ is
constant, and only for $E>E_{c}(L)$, where $p(E_{c}(L))=L^{-1}$, does
$p(E)$ gives the behavior of an infinite system. One way to obtain
$E_c$ is by extrapolating the part of the curve to where $p=0$. An
improved method is to extrapolate $E_{c}(L)$ to $L \rightarrow
\infty$ \cite{HB}. Eq. (\ref{Beqn}) guarantees that
$u_{j\alpha}$ and $v_{j\alpha}$ are either both extended or
localized, so it suffices to calculate $p(E)$ for $u_{j\alpha}$. In Fig. 4
we show $p(E)$ for various $L$'s in 1D for $\Delta = 1.5$ (SF side)
and $\Delta = 0.5$ (insulating side), which for the former seems to
show clearly a finite $E_c$. For the latter, we ascertain $E_c$ to be
very small, probably zero (our extrapolation actually gives an
unphysical small negative value).  This seems to support the idea that
the localization of the ground state and the excitations occur
simultaneously, a feature of Giamarchi and Schulz's theory in 1D \cite{GS}.

Upon reflection, however, we have serious doubts. In the
perturbative (in disorder) calculation of Ref. \cite{LZ}, the phonon
mean free path is found to diverge as $E^{-(d+1)}$. Common wisdom
has it that in 1D, the localization length and the mean free path are
essentially identical, since any scattering is backscattering. Hence,
one expects the relatively slow $\xi^{-1}(E) \propto E^{2}$ for small
$E$, crossing over to a more rapid $E$ dependence at a higher
cross-over energy $E_x$. This is in fact known to be the case for classical
vibrational modes in 1D \cite{ishii}, and all states except
the uniform translation mode are localized. The problem is
then that the divergence of the localization length $\xi$ is
not a true critical phenomena with a critical exponent. In our
calculation, if the size $L < \xi(E_{x})$, then we cannot probe
the weak $E$ dependent regime, and we will mistake $E_x$ for the
true mobility edge.  For comparison, we look at a finite system
of random masses connected by springs and find $p(E)$ curves
similar to Fig. 4. While our results do not constitute evidence for all
eigenstates of (\ref{hrrs}) to be localized in 1D, we believe this is in
fact the case, and what Fig. 4 shows is $E_x$ decreasing as the
disorder is increased, vanishing at or close to the superfluid-insulator
transition. Since even in 2D, it is believed that all classical waves are
localized \cite{JS}, it seems probable
that all phonon excitations are localized
too. This fact of localization of all the excitations, if it is true, may
invalidate the usual effective field theory based on the action of
propagating phase modes, which is crucial for the scaling theory of
Ref. \cite{FWGF}. Indeed, certain predictions of the scaling theory has
been questioned by recent quantum Monte Carlo simulations
\cite{MTU} on 2D hard-core dirty bosons.

In our model, the classical state is the Gutzwiller state, and $1/S$
serves as an expansion parameter for quantum fluctuations. We find
a transition at finite disorder when only gaussian fluctuations are
kept. Since the physics is sufficiently general, we believe this
conclusion would hold true if one consider soft-core (e.g., Hubbard
$U$) models and use $\hbar$ as the quantum expansion parameter.
What about a phase diagram of disorder vs. $U$? Since bosons will
condense into the lowest energy (localized) state for $U = 0$, the
critical disorder = 0. How about $U \rightarrow 0$? The problem is
that increasing $U$ both affects the classical condensate and
enhances quantum fluctuations. In fact, the Hartree solution becomes
extended with infinitesimal $U$. Thus, the $U \rightarrow 0$ limit
should not be qualitatively different from the limit of $1/S
\rightarrow 0$, and the critical disorder
is again finite. This conclusion is in agreement with numerical works
performed on the disordered boson Hubbard model \cite{MC}.

In 1D, Ref. \cite {GS} predicted that the renormalized critical exponent
$\eta $ is universal and equal to 1/3 at the SF-BG transition
based on a perturbative renormalization group calculation.  One might ask
what the value of $\eta$ is in the gaussian theory. Rigorously, the
spin wave theory as formulated cannot produce a power-decaying
correlation function (it is similar to using $(\nabla \pi)^2$ as the action
in the classical non-linear $\sigma$ model ). However, $\eta$ and
$\gamma$, the coefficient of $\ln N$ in $\delta m/m$, are proportional
in the pure case.  Assuming the relation holds even with disorder, it
implies
\begin{equation}
\frac{\eta_c}{\eta_{pure}} =\frac{\gamma_c}{\gamma_{pure}} \;\;.
\end{equation}
{}From the slope of the $\Delta =0.6$ curve in Fig. 1, we estimate
$\eta_{c}/\eta_{pure} \approx 1.4$. This ratio
is universal, while $\eta_c$ is not. $\eta_c \rightarrow 0$ as $1/S
\rightarrow 0$ in our model, or as $U \rightarrow 0$ in Hubbard
type models.

We thank Y. Huo, D.H. Lee, R. Pandit and K. Runge for helpful
comments.  MM acknowledges hospitality of the Aspen Center for
Physics;  LZ acknowledges support by the National Science
Foundation under Grant No. DMR-9101542 and by the U.S.
Department of Energy through Contract No. DE-AC05-84OR21400
administered by Martin Marietta Energy Systems Inc..

\begin{figure}
\caption{
Fluctuation corrections to the order parameter $\delta m/m$
is plotted against $\ln N$. The divergence becomes faster than $\ln N$
as $\Delta $ exceeds a critical value $\Delta_{c} \approx 0.6$.
The solid lines are obtained through a linear fit and the dashed line
is a guide to the eyes.
}
\end{figure}

\begin{figure}
\caption{
The same as Fig. 1, plotted for 2D systems. Unlike it in 1D,
$\delta m$ is finite for weak disorder and diverges for $\Delta >
\Delta_c$, which is between 0.1 and 0.08.  ($b$) is the same plot
as it in ($a$), but is presented on a different scale,  which
shows clearly that $\delta m$ is bounded as $N \rightarrow \infty$
for small $\Delta$.
}
\end{figure}

\begin{figure}
\caption{
DOS on the insulating side ($\Delta = 0.5$) in 1D. Here $N=100$.
}
\end{figure}

\begin{figure}
\caption{
Participation ratio $p(E)$ in 1D is plotted for several values
of $N$. The pseudo-mobility edge $E_{c}$ is obtained from these
plots through the procedure described in the main text.
In ($a$), $\Delta = 1.5$, the system is in the superfluid
phase and $E_{c}$ is finite. $E_{c}$ is about zero in the
insulating phase as it shows in ($b$) for $\Delta = 0.5$.
}
\end{figure}


\begin{references}
\bibitem{HM}
Our philosophy here deviates from the recent study by
Huang and Meng (Phys. Rev. Lett., {\bf 69}, 644 (1992)), who
look for the transition by considering $\delta\rho_{s} = - \rho_s$.
\end{references}
\end{document}